# A new model of correlated disorder in relaxor ferroelectrics

A. Bosak[1], D. Chernyshov[2], S. Vakhrushev[3], M. Krisch[1]

[1]European Synchrotron Radiation Facility, BP 220, 38043 Grenoble Cedex, France

[2]Swiss-Norwegian Beam Lines, BP 220, 38043 Grenoble Cedex, France

[3]Ioffe Physico-Technical Institute, St. Petersburg, 194021, Russia

We propose a simple phenomenological picture to explain the unusual dielectric properties of the proto-typical ferroelectric relaxor lead magnesium niobate-titanate (PMN-PT). Our model assumes a specific, slowly changing, displacement pattern of the lead ion, which is indirectly controlled by the low-energy acoustic phonons of the system. The model qualitatively explains in great detail the temperature, pressure, and electric field dependence of diffuse neutron- and x-ray scattering, as well as the existence of hierarchy in the relaxation times of these materials. We furthermore show that the widely used concept of polar nanoregions as individual static entities is incompatible with the available body of experimental diffuse scattering results.

Relaxor ferroelectrics are known since more than fifty years[1] and have attracted significant interest because of their numerous unusual properties such as the appearance of a broad peak in the real part of the dielectric permittivity as a function of temperature. This peak decreases in magnitude and shifts to higher temperature with increasing probe frequency over a very large frequency domain[2]. They also have a variety of applications as sonar projectors for submarines and surface vessels as well as sensors and actuators. One of the most widely used concepts in relaxor physics is the model of polar nanoregions (PNR), first proposed in 1983[3]. The starting model is based on the existence of small regions (down to few unit cells in some dimensions) of local polarization, with the polarization parallel to specific high-symmetry directions. It is commonly accepted that the strong neutron- and x-ray diffuse scattering in the vicinity of Bragg reflections is a signature of PNRs[4], and several microscopic models have been evoked: the classical "pancake" model[5], the generalized pancake model[6], interdomain atomic shifts[7], correlated atomic displacements[8], and anisotropic strain[9]. These models either give reasonable agreement only for a limited set of directions/planes in reciprocal space and fail elsewhere, or predict features which are not observed experimentally. As an example, none of the reported models does predict the local minima of intensity appearing in the direction, roughly parallel to the momentum transfer of the corresponding Bragg node.

Here, we present a detailed synchrotron x-ray diffuse scattering study of lead magnesium niobate-titanate $PbMg_{1/3}Mb_{2/3}O_3$-$PbTiO_3$ (PMN-PT), a prototypical ferroelectric relaxor. We perform a semi-quantitative analysis of the scattering pattern, which reproduces all experimental features remarkably well, and moreover provides strong evidence that in this system and the related family of ferroelectric compounds such as $PbZn_{1/3}Nb_{2/3}O_3$-PT and $PbMg_{1/3}Ta_{2/3}O_3$ static polar nanoregions do not exist as individual entities.

The PMN-PT crystal with a Ti content of 6 at. % in the B-sublattice was grown using a top-seeded solution growth technique[10]. For the measurements a needle-like fragment of ~50 μm thickness was utilized. Its surface was etched with hot concentrated hydrochloric acid, and the experiment was performed at room temperature. The diffuse scattering dataset was collected at the Swiss-Norwegian Beam Lines at ESRF with a MAR345 image plate detector at a wavelength of 0.700Å with an angular step of $0.2^0$. The experimental geometry was refined using the CrysAlis software[11]; and the resulting parameters were used for the 3D reciprocal space reconstruction. Some of the cubic symmetry elements were applied in order to increase the signal-to-noise ratio. For visualization purposes USCF Chimera[12] and Pov-Ray[13] packages were used.

The left parts of Fig. 1a and 1b show the experimental reciprocal space cuts of the HK0 and HKK planes for PMN-PT. The diffuse scattering, observed in the proximity of Bragg reflections, is a sum of three components: i) an intense relaxor-specific contribution (butterfly-shaped around *H00* and ellipsoid-like around *HH0*); ii) Huang scattering due to local distortions of the lattice, iii) thermal diffuse scattering (TDS) due to phonons. The specific shape of Huang scattering can be easily identified at high temperature[14], where the relaxor-specific component disappears, and the weaker TDS component remains practically hidden underneath the other two components. The observed diffuse scattering in the present case is therefore dominated by the relaxor-specific contribution.

We first note that diffuse spots with non-integer Miller indices visible in Fig. 1b point towards correlated Mg/Nb disorder, where nearest neighbors tend to be of different type. We further note that the observed distribution of diffuse intensity strongly resembles the pattern of thermal diffuse scattering for cubic crystals; while it is clear from the above said that in the present case the diffuse scattering is essentially of (quasi)elastic nature[15].

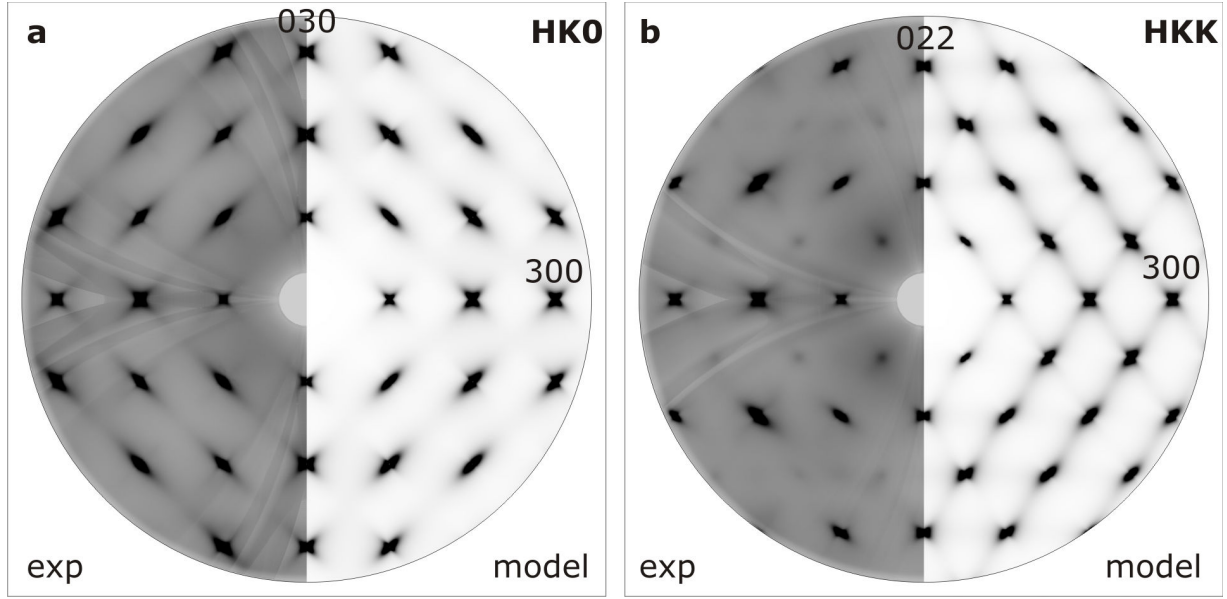

FIG. 1. Experimental and model diffuse scattering intensity distribution of PMN-PT in (a) HK0 and (b) HKK planes. For the reconstruction of the experimental patterns the Laue symmetry of crystal was applied. Spots with non-integer indices should be disregarded (see text). The observed extra V-shaped intensity variations are artefacts of the reconstruction.

This observation led us to model the diffuse scattering in close analogy to the formalism for TDS:

$$I \propto f_{Pb}^2(\mathbf{Q})\exp(-2W_{Pb}(\mathbf{Q}))\frac{\sin^2(2\pi r_0 Q)}{Q^2} \cdot \mathbf{Q}^T \cdot \frac{\coth\left(\alpha\sqrt{\Re(\mathbf{Q})}\right)}{\sqrt{\Re(\mathbf{Q})}} \cdot \mathbf{Q} \qquad (1)$$

where Q – momentum transfer, $f_{Pb}$ – atomic scattering factor of the lead ion[16], $W_{Pb}$ – Debye-Waller factor[17], and $\alpha$ – constant for a given temperature. The $\sin(2\pi r_0 Q)/Q$ term describes the possible locations of the Pb ion over a shell of radius $r_0$[17], and the symmetric tensor $\Re(\mathbf{Q})$ is defined as:

$$\Re_{11}(\mathbf{Q}) = \Psi_{11}(2 - \cos(2\pi Q_x)(\cos(2\pi Q_y) + \cos(2\pi Q_z))) + (2\Psi_{44} - \Psi_{12})(1 - \cos(2\pi Q_y)\cos(2\pi Q_z))$$
$$\Re_{12}(\mathbf{Q}) = (\Psi_{44} + \Psi_{12})\sin(2\pi Q_x)\sin(2\pi Q_y) \qquad (2)$$

where the other components of the tensor can be deduced by cyclic permutations.

Near reciprocal lattice nodes $\boldsymbol{\tau}$ ($\mathbf{Q} = \boldsymbol{\tau} + \mathbf{q}$) $\Re(\mathbf{Q})$ transforms to $\Re_{jk}(\mathbf{q}) = 4\pi^2 \Psi_{ijkl} q_i q_l$. Here, the tensor $\Psi$ accommodates cubic symmetry, and consequently we obtain in Voigt notation the following equation:

$$\Re_{jk}(\mathbf{q}) = 4\pi^2 \begin{pmatrix} \Psi_{11} q_x^2 + \Psi_{44}(q_y^2 + q_z^2) & (\Psi_{12} + \Psi_{44}) q_x q_y & (\Psi_{12} + \Psi_{44}) q_x q_z \\ (\Psi_{12} + \Psi_{44}) q_x q_y & \Psi_{11} q_y^2 + \Psi_{44}(q_x^2 + q_z^2) & (\Psi_{12} + \Psi_{44}) q_y q_z \\ (\Psi_{12} + \Psi_{44}) q_x q_z & (\Psi_{12} + \Psi_{44}) q_y q & \Psi_{11} q_z^2 + \Psi_{44}(q_x^2 + q_y^2) \end{pmatrix} \quad (3)$$

We again emphasize that these expressions closely resemble the ones for the description of thermal diffuse scattering[18], but here $\Psi_{ij}$ does not have the meaning of an elastic modulus, and $\alpha$ does not have the meaning of an inverse temperature. Besides scaling by an overall intensity factor, our model depends on only three parameters: $\Psi_{ij}\alpha^2$; moreover, in the very proximity of reciprocal lattice nodes this number reduces to two, for example $\Psi_{12}/\Psi_{11}$ and $\Psi_{44}/\Psi_{11}$. The structure of $\Re(\mathbf{Q})$ as given by Eq. 2 creates extra Bragg spots, corresponding to face-centering in real space. These artefacts are situated far away from the region of interest, and therefore do not affect our analysis and interpretation. In fact, our proposed description remains valid for a larger region in the proximity of nodes than a similar model mimicking the primitive cubic lattice. Taking the experimental values for W(Q) and $r_0 = 0.286$A[17], and adjusting the free parameters to $\Psi_{11}\alpha^2 = 0.121$, $\Psi_{12}\alpha^2 = 0.097$ and $\Psi_{44}\alpha^2 = 0.049$, we reproduce extremely well both 2D (Fig. 1) and 3D diffuse scattering patterns (Fig. 2) in the proximity of Bragg reflections both in terms of shape and relative intensities up to at least 0.1-0.15 r.l.u.. The proposed parameterization must fail both at very small q, where the system behaves like a continuum, and for large q, where the approximation of the perovskite lattice dynamics by a fake monoatomic lattice is too simplistic. The range of validity of the model can be extended at small q limit by introduction of an additive isotropic tensor term in the Eqs. 1-3. Such an extension would account for $\frac{1}{q^2 + k^2}$ dependence of diffuse scattering, experimentally observed at small $q$, with $k$ being a correlation radius[19]. Future development of the model should also take into account the temperature dependence of the related correlation characteristics.

The origin of this term, introduced on the phenomenological basis, can be linked to a variety of microscopic mechanisms. One of the possible mechanisms relates to the fact that the ionic displacements corresponding to the observed diffuse scattering contain both acoustic-like (deformation) and optic-like (polarization) components[20]. For perovskites acoustic and optic modes at $q = 0$ can be parameterized jointly as belonging to the same irreducible representation[21]. By the analogy, similar parameterization can be used to introduce $q^2 + k^2$ eigenvalues dependence to Eqs. 2-3.

Second possibility links to the finding that in a media with large concentration of point defects elastic interactions may be screened similar to Debye screening of electrostatic potential[22]. Relaxors are intrinsically very defect crystals, since every unit cell is distorted, first of all, due to omnipresent displacement of lead ions from the average position. Such kind of screening will contribute to Eqs. 2-3 in the same way as in previous case, but now with $k$ being a screening constant.

However, at the moment we would prefer to leave the details of the low-$q$ behavior for the future analysis and concentrate only at the intermediate $q$-range where only 3 or 2 parameters provide perfect description.

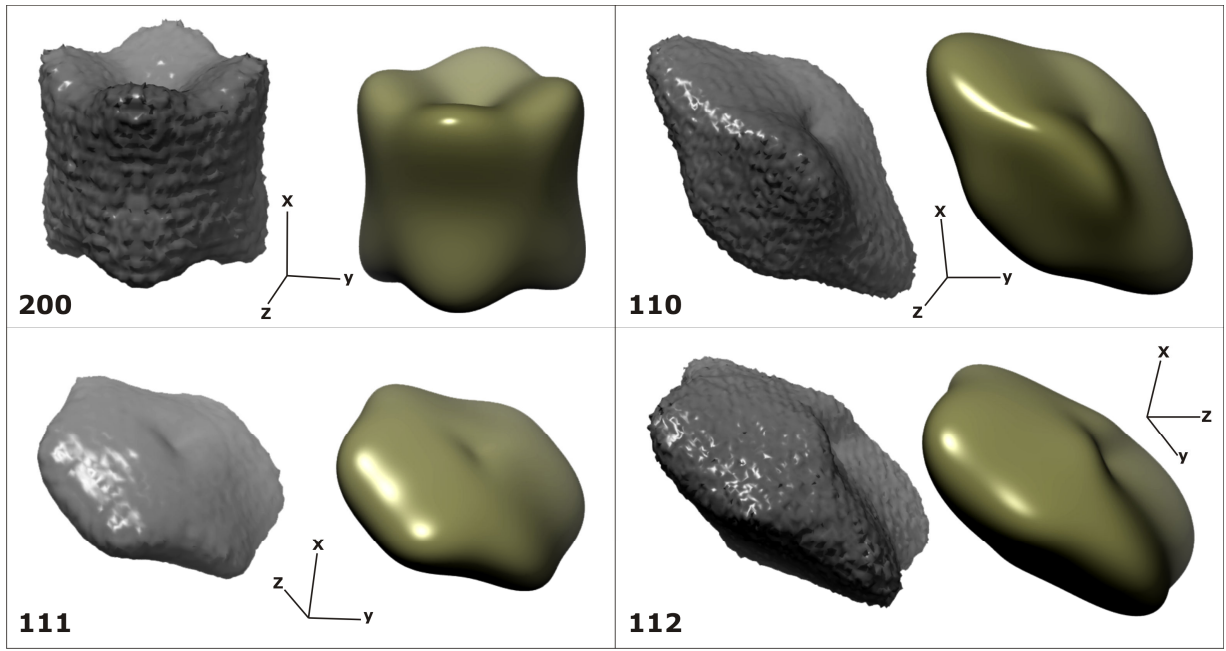

FIG. 2. Comparison of experimental (PMN-PT) and model isosurfaces of diffuse scattering intensity in the proximity of various Bragg nodes.

Based on the excellent agreement between experiment and model, and in the light of previous temperature, pressure and electric field dependent diffuse scattering studies in PMN-PT or related compounds, we propose the following mechanistic concept. In each unit cell the Pb ions are displaced from their 1a positions. The set of possible positions is very large and can be approximated by a spherical shell[17]. Under the assumption that neighboring Pb ions do not substantially interact between each other but only *via* the $BO_3$ octahedral framework, they can be aligned together by a dynamical local distortion: a wave (phonon) with sufficient amplitude and proper polarization. The probability to obtain a distortion of sufficiently large amplitude is inversely proportional to the frequency of the wave ($\sim \omega^{-2}$ for $\hbar\omega \ll k_BT$), and the Pb displacement pattern remains frozen on the scale of typical phonon lifetimes, unless it is affected by another wave/phonon. Consequently, the quasielastic diffuse scattering in first approximation mimics the thermal diffuse scattering pattern from acoustic phonons. Its shape will not necessarily coincide with the shape of thermal diffuse scattering as the interaction between the acoustic wave and the Pb displacement is not isotropic, but the symmetry of the underlying $\Psi$ tensor near Bragg nodes must be the same as for the elastic tensor. As relaxor-specific diffuse scattering is related to *large* displacements of Pb from its average position, its intensity must be superior to Huang scattering and TDS at least when it is seen by X-ray scattering. Further support for the validity of our model is provided by the fact that relative intensities of spots are reproduced in a satisfactory way using only Pb-related prefactor. Apparently, to account for neutron diffuse scattering where scattering factors for O and Pb are comparable, oxygen displacements may also become important[20,23]; an extension of the model incorporating the displacement of oxygen and also of Mg/Nb atoms implies mainly the modification of the prefactor in Eq. 1. However, such an extension will not affect strongly the local shape of diffuse scattering since the prefactor, irrespective on the structural motif involved in the calculation, is slowly varying function of **Q**.

The distribution of atomic displacements in space is therefore not static, but stationary in the sense of its power spectrum/diffraction and changing slowly in time compared to conventional lattice dynamics. In other words, the Pb motion can be considered as equivalent to a very low energy overdamped strongly anharmonic mode. It is not clear, whether – and how – this mode can be directly observed, but the corollaries of this model can be corroborated by the following well known experimental facts. i) The disappearance of a relaxor-specific component of diffuse scattering at high temperature[14] can be explained by the transition of Pb displacements towards a free uncorrelated movement over the spherical shell. As a consequence, the structured diffuse scattering transforms to a smoothly changing background. ii) The reduction of diffuse scattering due to the Pb movements under high pressure[24] follows from the unavoidable creation of deep local minima on the spherical shell, in which the Pb ions remain frozen. iii) Changes of the diffuse scattering in an applied electric field[25] correspond to the creation of additional anisotropy in the energy relief over the displacement shell. As a result, diffuse scattering features perpendicular to the field direction should shrink – in agreement with experimental observations; moreover, further increase of the electric field results in the complete suppression of the diffuse signal[26]. iv) The extremely large spread of relaxation times[2] can be associated with a hierarchy of displacement patterns in space and their respective lifetimes.

Despite the simplicity of our model - with only three adjustable parameters - it provides the best description of the complex relaxor-specific diffuse scattering ever reported. It is important to emphasize that a quantitative derivation of the relevant parameters is still missing. The fact that very complex diffuse pattern may be efficiently reduced to only three numbers should stimulate the further development of new phenomenological models capturing the essential physics of relaxors. However, for a complete theoretical understanding of the underlying mechanisms, molecular dynamics simulations may become the key tool.

**Acknowledgments**

We are grateful to Efim Kats (Institut Laue-Langevin, Grenoble, France) for the numerous fruitful discussions and the encouragement.